\documentclass[showpacs,twocolumn,prb]{revtex4}
\usepackage{graphicx}

\newcommand{\be}{\begin{equation}}
\newcommand{\ee}{\end{equation}}
\newcommand{\bea}{\begin{eqnarray}}
\newcommand{\eea}{\end{eqnarray}}
\newcommand{\bwt}{\begin{widetext}}
\newcommand{\ewt}{\end{widetext}}

\begin{document}

\title{Ground state properties of the Holstein model near the adiabatic limit}

\author{Zhou Li, D. Baillie$^1$, C. Blois$^2$, F. Marsiglio}
\affiliation{Department of Physics, University of Alberta, Edmonton, Alberta, Canada,
T6G~2G7}

\begin{abstract}
We adapt a variational procedure to calculate ground state properties of the Holstein
model in the adiabatic limit. At strong coupling, this adaption leads to rapid
convergence of results. The intermediate coupling regime is further handled with an
adaptive algorithm. We also use semi-classically derived results for the adiabatic
end-point, along with weak coupling perturbation theory. These establish weak and
strong coupling (or large and small polaron, respectively) regimes in two dimensions
or higher. As is well known, these are connected smoothly, but the cross-over becomes
increasingly abrupt as the phonon frequency decreases.
\end{abstract}

\date{\today }
\maketitle

\section{introduction}

There has been considerable work performed over the last two decades on
the Holstein model \cite{holstein59}. Interest in this model is fueled by
the fact that it serves as the paradigm for electron-phonon interactions, much
like the Hubbard model \cite{hubbard63} serves the same purpose for electron-electron interactions.
While a considerable amount of this work has focussed on the many-electron problem, another subset
has examined the single-electron, or polaron problem. A recent review of this work is available,
for example, in Ref. [\onlinecite{fehske07,alexandrov07}].

In our opinion, the most promising numerical technique for determining polaron properties in the
thermodynamic limit is the variational procedure outlined by Trugman and coworkers \cite{trugman90, bonca99}.
With this method properties such as the ground state energy and the effective mass are readily obtained,
in any dimension, over almost all parameter regimes. One range of parameter space that has remained
difficult, however, is near the adiabatic limit, which is what we address in this paper. The actual adiabatic
limit was first treated by Kabanov and Mashtakov \cite{kabanov93}; they found that in one dimension (1D), the electron
retains polaronic character for {\it all} electron-phonon coupling strengths, while in two dimensions and
higher there is a critical coupling strength, below which the electron behaves in a `free-electron-like'
manner, and above which it is polaronic. At the same time, away from the adiabatic limit the problem is
known from numerical solutions to have a smooth crossover as a function of coupling strength (i.e. no
abrupt transition), so it is of interest to pursue this crossover as the phonon frequency decreases towards
zero. This was done to some degree in Refs. \onlinecite{alexandrov94,marsiglio95}, but only for rather small
lattices in one dimension. Our aim is to examine this limit using the Trugman variational technique
\cite{bonca99,ku02}.

The outline of the paper is as follows. In the next section we outline the model,
and establish notation, etc. In Section III we describe some refinements to the
variational method, and provide some illustrative examples to demonstrate the
improvement in convergence. In Section IV we provide some numerical results as the adiabatic
parameter $\omega_E/t$ approaches zero. Also provided are some perturbation theory
results \cite{marsiglio95}, which can be reinterpreted to provide constraints for the
numerical results. In Section V we show some results concerning the expected numbers
of phonons in the ground state, which gives another indication of the difficulty of the
adiabatic limit. Finally, we close with a summary.

\section{the model}

The model that most simply describes an electron interacting with optical
phonons is the Holstein Hamiltonian, given by
\begin{eqnarray}
H &=& -t\sum_{i,\delta} \bigl(c_i^\dagger c_{i+\delta} +
c_{i+\delta}^\dagger c_{i}\bigr) \nonumber \\
&+& \sum_i \bigl[ {p_i^2 \over 2M} + {1 \over 2}M\omega_E^2 x_i^2\bigr] - \alpha \sum_{i} n_i x_i
\label{ham_x}
\end{eqnarray}
which is also written
\begin{eqnarray}
H &=& -t\sum_{i,\delta} \bigl(c_i^\dagger c_{i+\delta} +
c_{i+\delta}^\dagger c_{i}\bigr) \ - \ g\omega_E \sum_{i} n_i \bigl(a_i +
a_i^\dagger\bigl)  \nonumber \\
&+& \omega_E \sum_i a_i^\dagger a_i,  \label{ham}
\end{eqnarray}
where $c_i^\dagger$ ($c_i$) creates (annihilates) an electron at site $i$
(the spin label is suppressed) and $n_i \equiv c_i^\dagger c_i$ is the number density
operator. The ion momentum $p_i$, and displacement $x_i$ are quantized via
\begin{eqnarray}
x_i &\equiv& \sqrt{1 \over 2M\omega_E} \bigl( a_i^\dagger + a_i \bigr) \nonumber \\
p_i &\equiv& i\sqrt{M\omega_E \over 2} \bigl( a_i^\dagger - a_i \bigr),
\label{a_defns}
\end{eqnarray}
where $M$ is the ion mass (we set $\hbar \equiv 1$) and $a_i^\dagger$ ($a_i$) creates
(annihilates) a phonon at site $i$. The sum over $i$ is over all sites in
the lattice, whereas the sum over $\delta$ is over nearest neighbors. Here,
as the notation already suggests, we confine ourselves to nearest neighbor
hopping only. The parameters are the hopping integral $t$, the phonon
frequency $\omega_E$, and the coupling of the electron to the oscillator
degrees of freedom, $\alpha$. This parameter is the bare coupling between the electron
and the ion; however, it is rarely used, and instead in the polaron literature the dimensionless
coupling constant $g \equiv {1 \over \omega_E}{ \alpha \over \sqrt{2M\omega_E}}$ is used. In the many-body literature, the dimensionless parameter $\lambda \equiv {2\omega_E g^2 \over W} \equiv {\alpha^2 \over M\omega_E^2 W}$ is used, where $W \equiv 2zt$ is the electronic bandwidth for a cubic
tight-binding model with coordination number $z$ ($z=2,4,6$ in $1, 2, 3$
dimensions, respectively). The parameter $\lambda$ has historical
significance for the effective mass of degenerate electrons weakly coupled
to phonons. Alternatively, and most useful in the strong coupling regime,
the parameter $g$ (or $g^2$) was used in the Lang-Firsov transformation \cite{lang63}, and
leads to a band narrowing factor $t \rightarrow t^\ast = t e^{-g^2}$ in
first order degenerate perturbation theory. We write all energy scales in terms of the
hopping integral, $t$, which, hereafter is set to unity. So two dimensionless
parameters which usefully characterize this problem \cite{fehske07} are $\omega_E/t$,
the adiabaticity parameter, and, $\lambda \equiv 2g^2\omega_E/W$. Actually, as we argue
below in Section III, in two dimensions it is arguably more useful to use
$\lambda \equiv 2g^2 \omega_E/(4\pi t) = g^2 \omega_E/(2 \pi t)$, where $1/(4\pi t)$
is the value of the non-interacting electron density of states at the bottom of the band (as
opposed to the average density of states).

As mentioned in the introduction this model has been most successfully
analyzed using a refinement of the standard Lanczos method due to Trugman
\cite{trugman90,bonca99}. Very accurate results can be obtained in any
dimension \cite{ku02} in almost all parameter regimes \cite{bonca99,ku02,fehske07}.
A difficulty remains for moderately to strongly
coupled systems with low adiabaticity parameter $\omega_E/t$. For example,
if one uses the Lang-Firsov transformation \cite{lang63} to define the
zeroth order strongly coupled wave function, then the average number of
phonons in the ground state can be readily determined to be approximately $%
g^2$. For typical parameters in the moderately coupled regime (in one
dimension), say $\omega_E = 0.05t$, and $\lambda = 1.0$, then $g^2 = 40$,
and this is the approximate number of phonons in the ground state. The
Trugman procedure starts with a bare electron; on a moderate work station a
feasible number of applications of the Hamiltonian is $N_h = 22$ (as
remarked in Ref. \onlinecite{bonca99}), which produces a Hilbert space of
order $10^7$. This process with $N_h = 22$ produces states that contain a
maximum of 22 phonons, and cannot possibly yield the correct ground state.

\section{refinement of the Trugman method}

\begin{figure}[tp]
\begin{center}
\includegraphics[height=2.7in,width=2.7in]{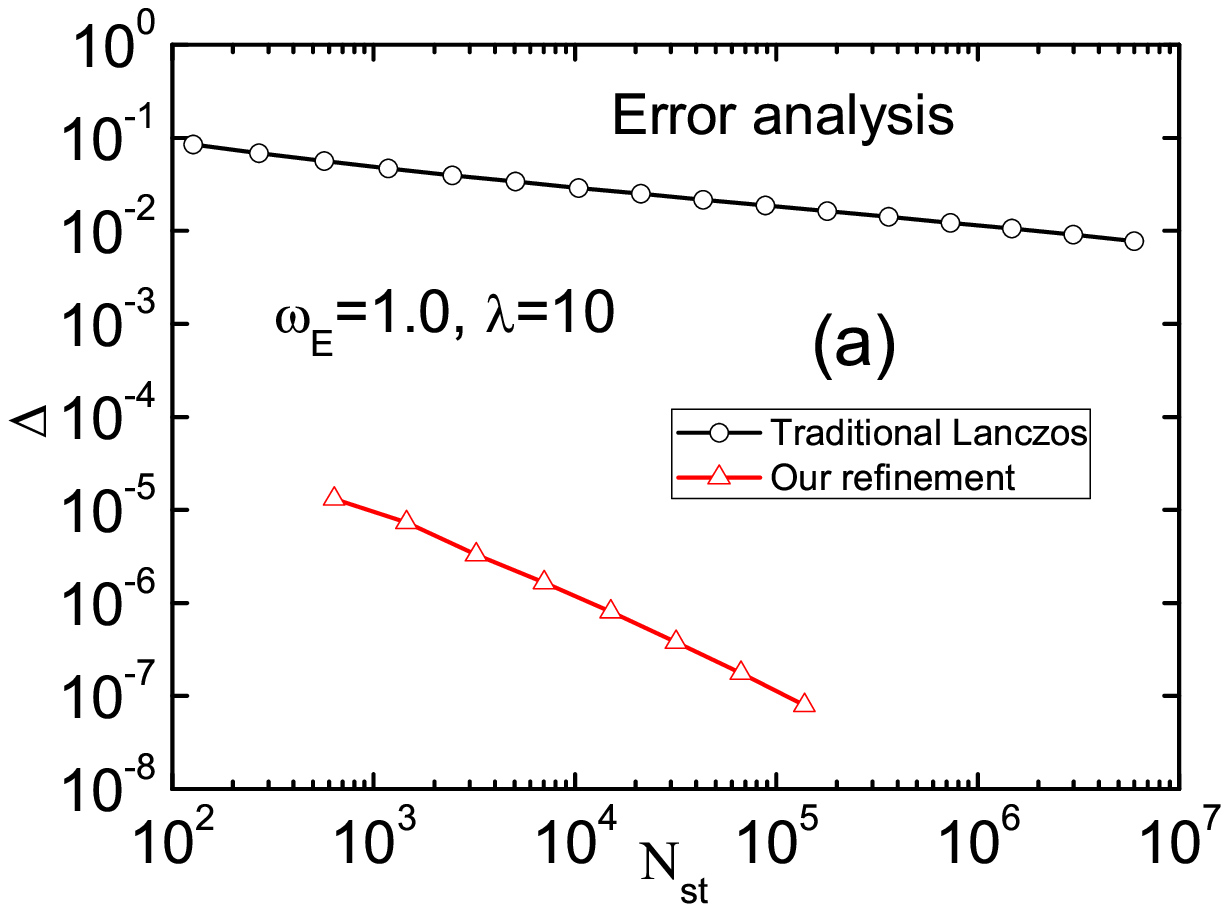}
\includegraphics[height=2.7in,width=2.7in]{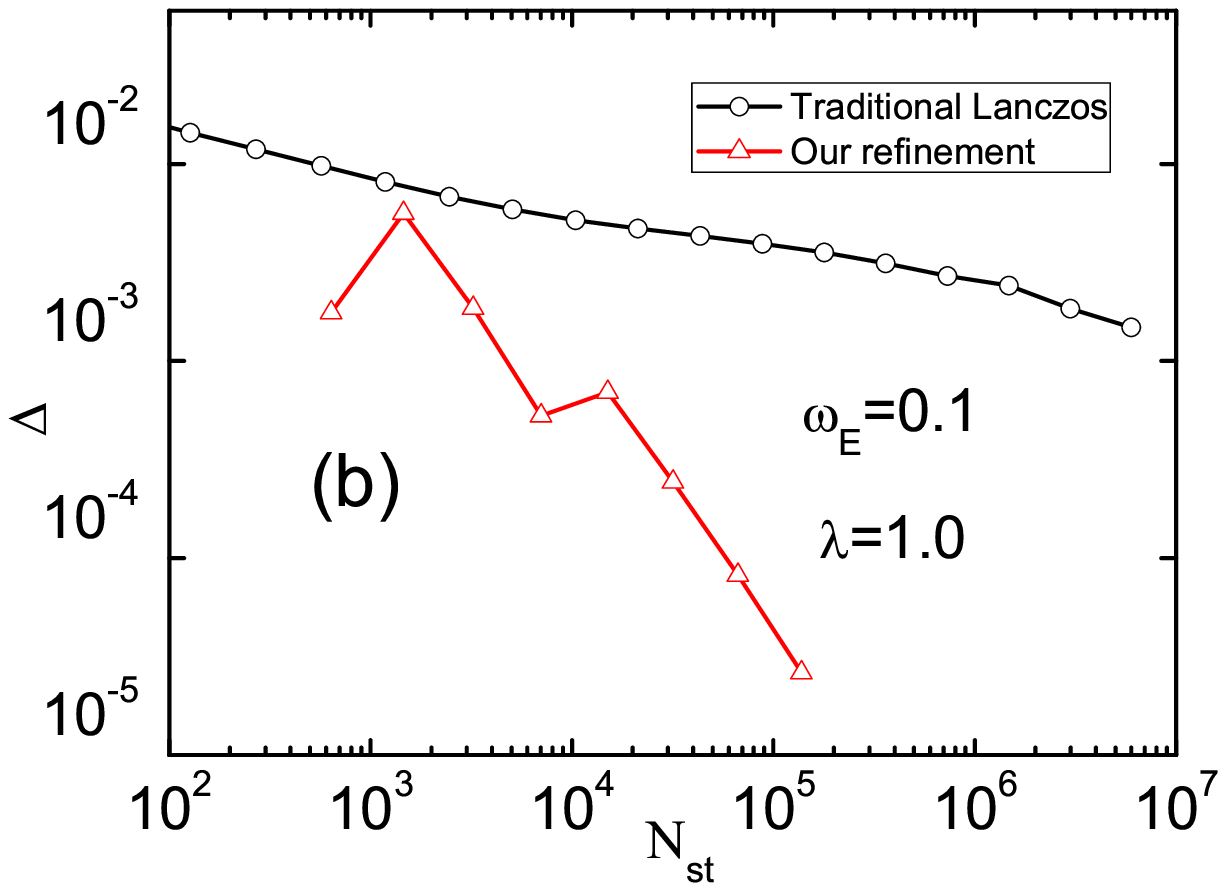}
\includegraphics[height=2.7in,width=2.7in]{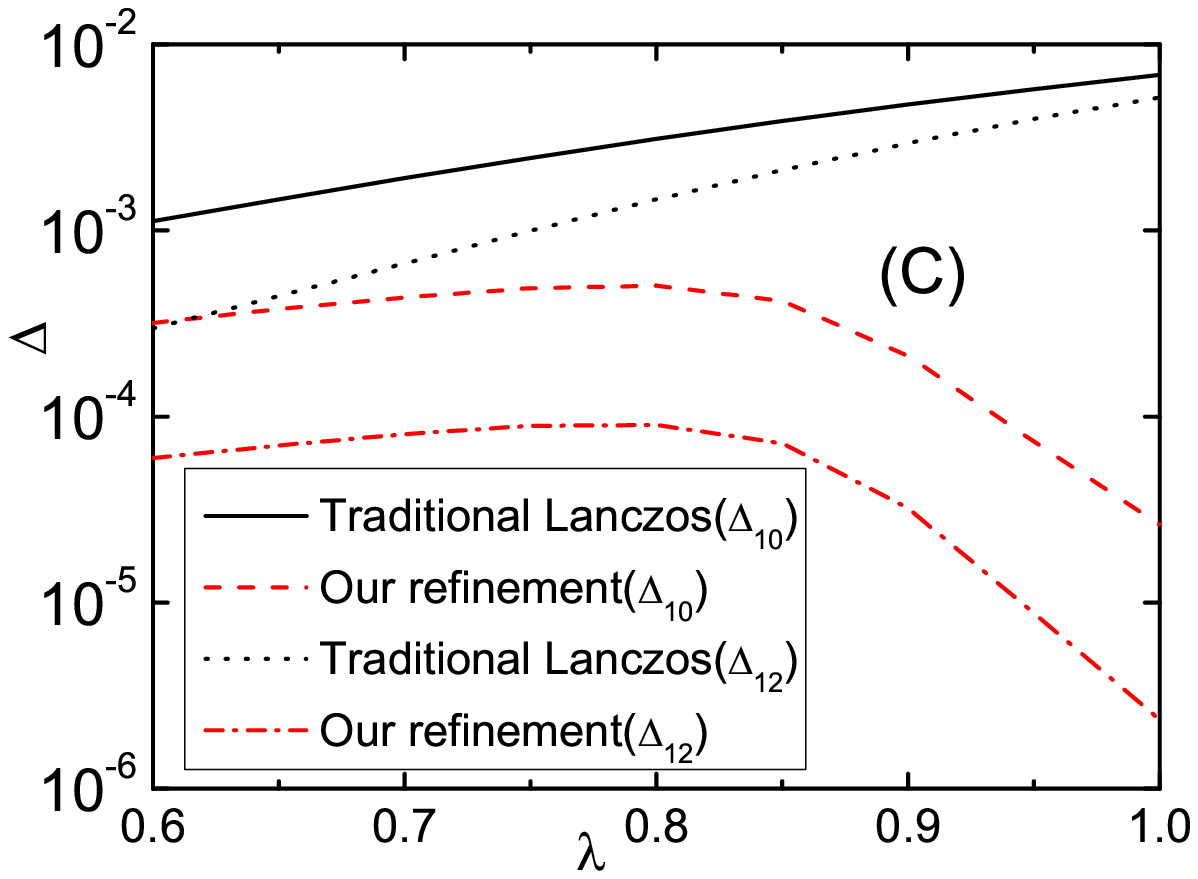}
\end{center}
\caption{ (color online) Fractional error vs. number of states for (a) a
very strong coupling case, and (b) a moderate coupling, but with $\omega_E/t = 0.1$.
In both cases our refinement speeds up convergence considerably. In (c) we show
the improvement over a range of coupling strength near $\lambda = 1$; for $\lambda >1$
(not shown) it is clear that the refinement leads to much better convergence. Note that
the fractional error does not have to decrease monotonically with the number of states added.}
\label{fig1}
\end{figure}

We have examined two simple refinements to the Trugman method \cite{li09};
instead of starting with the bare electron state (properly
extended throughout an infinite lattice), we first start with the state
which is used as the unperturbed state in the strong-coupling limit \cite{lang63,marsiglio95}:
\begin{equation}
|\psi \rangle = e^{-g^2/2} \sum_\ell e^{ikR_i} e^{-g\hat{a}_\ell^\dagger} \hat{c}%
_\ell^\dagger |0 \rangle,  \label{sc_zero}
\end{equation}
where the sum is over all lattice sites\cite{comment1,alvermann10}. As we shall see in
what follows this speeds up convergence considerably in the strong coupling
regime (either $\lambda >> 1$ or $\omega_E <<t$). An example of the
increased convergence is illustrated in Fig. \ref{fig1}. 
Following Fig. 2 of Ref. \onlinecite{ku02} we show the fractional error
$\Delta_N \equiv |(E_N - E_{N-1})/E_N|$ as a function of the number of states kept in the Hilbert space, for
two sets of parameters, both of which have $g^2 = 20$, using the bare
electron as the starting state and using the strong-coupling solution as the
starting state. There is a clear numerical advantage to using the latter. In
Fig. \ref{fig1}(c) we show the fractional error for a parameter regime near $%
\lambda \approx 1$, where the strong coupling start is better even for
values of $\lambda<1$. It is also clear that as $\lambda$ increases beyond
the range of this figure, the refinement becomes increasingly useful.

In pursuit of more severe disparate electron and phonon energy scales we
found that even starting with the strong coupling solution resulted in slow
convergence when $\lambda$ was of order unity. A remedy to this difficulty
is the following procedure: start at large values of $\lambda$, where
convergence is readily obtained after a few iterations. Lower the value of $%
\lambda$ by a small amount, and use as a starting wave function the previous
solution, truncated to include components with some minimal amplitude (so
that a few hundred basis states at most are used). Then converge the
solution for this value of $\lambda$, lower it, and continue the process
until the desired range is covered. We have found this latter procedure to
be the most robust, particularly when the phonon frequency is much smaller
than the electron hopping parameter, $t$. 
\begin{figure}[tp]
\begin{center}
\includegraphics[height=3.0in,width=3.0in]{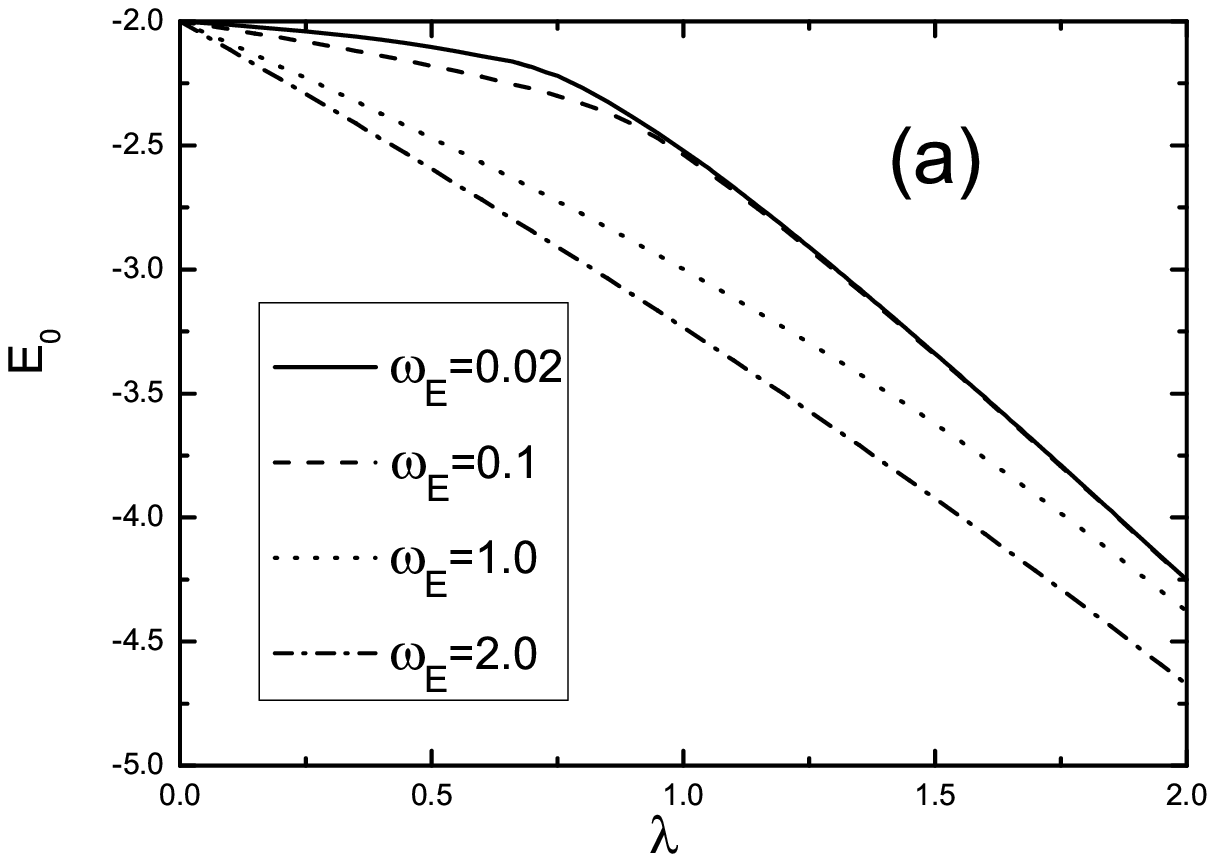}
\includegraphics[height=3.0in,width=3.0in]{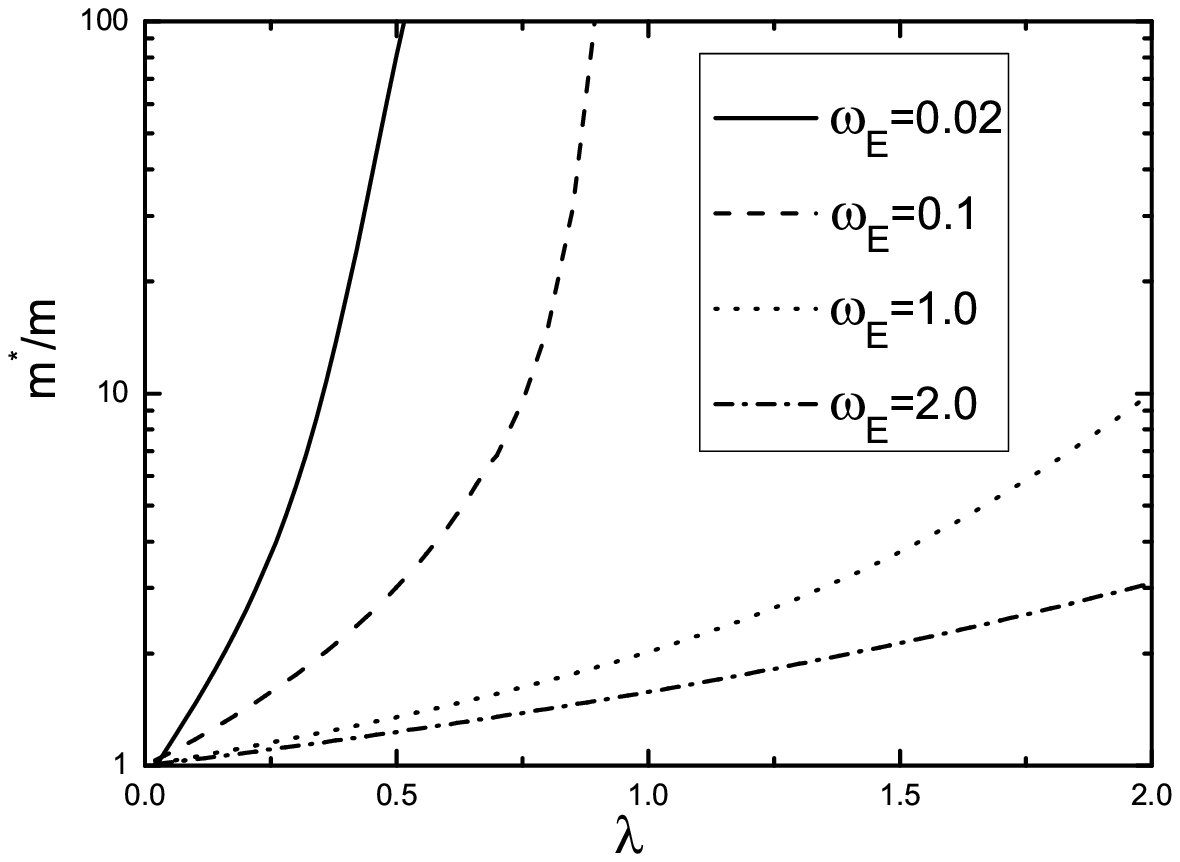}
\end{center}
\caption{ (color online) Ground state energy $E_0$ vs. $\lambda$ and $m^\ast/m$
vs. $\lambda$ for various phonon frequencies, in one dimension. There really is
no special value of $\lambda$ singled out in these curves, consistent with the
crossover phenomenon discussed in the text.}
\label{fig2}
\end{figure}
%
\begin{figure}[tp]
\begin{center}
\includegraphics[height=3.0in,width=3.0in]{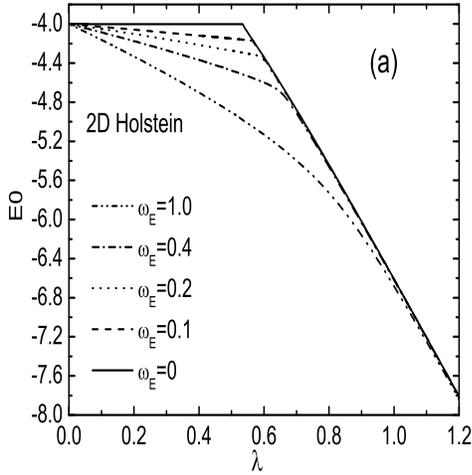}
\includegraphics[height=3.0in,width=3.0in]{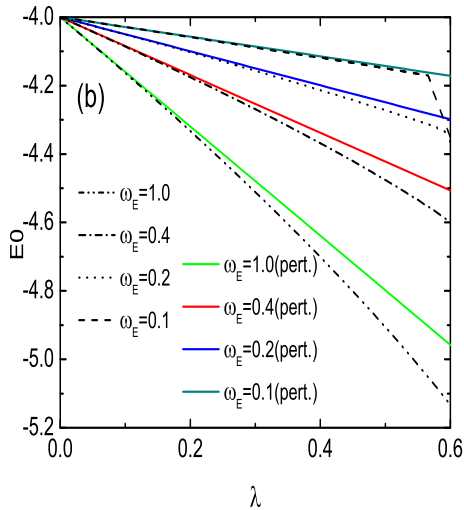}
\end{center}
\caption{ (color online) (a) Ground state energy $E_0$ vs. $\lambda$ for various
phonon frequencies, in two dimensions. In contrast to the 1D results in
Fig. \protect\ref{fig2}, a `special' value of
$\lambda$ is now apparent --- $\lambda \approx 0.55$. However, for any non-zero
phonon frequency the behaviour below and above this special value is smoothly
connected. Only in the adiabatic limit does the behaviour change abruptly.\\
(b) Expansion of the weak coupling regime showing the numerical results along side
the perturbation theory results. Agreement is very good.}
\label{fig3}
\end{figure}

\begin{figure}[tp]
\begin{center}
\includegraphics[height=3.0in,width=3.0in]{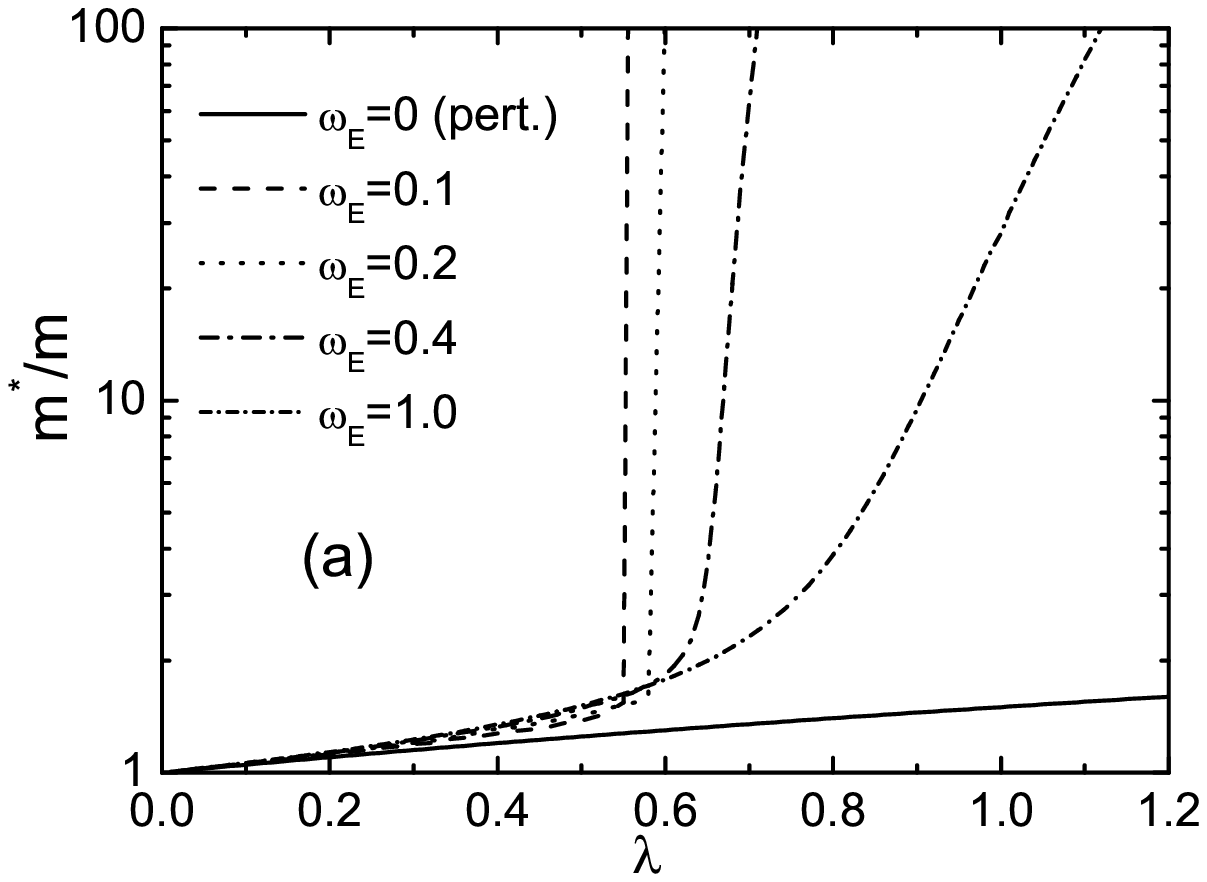}
\includegraphics[height=3.0in,width=3.0in]{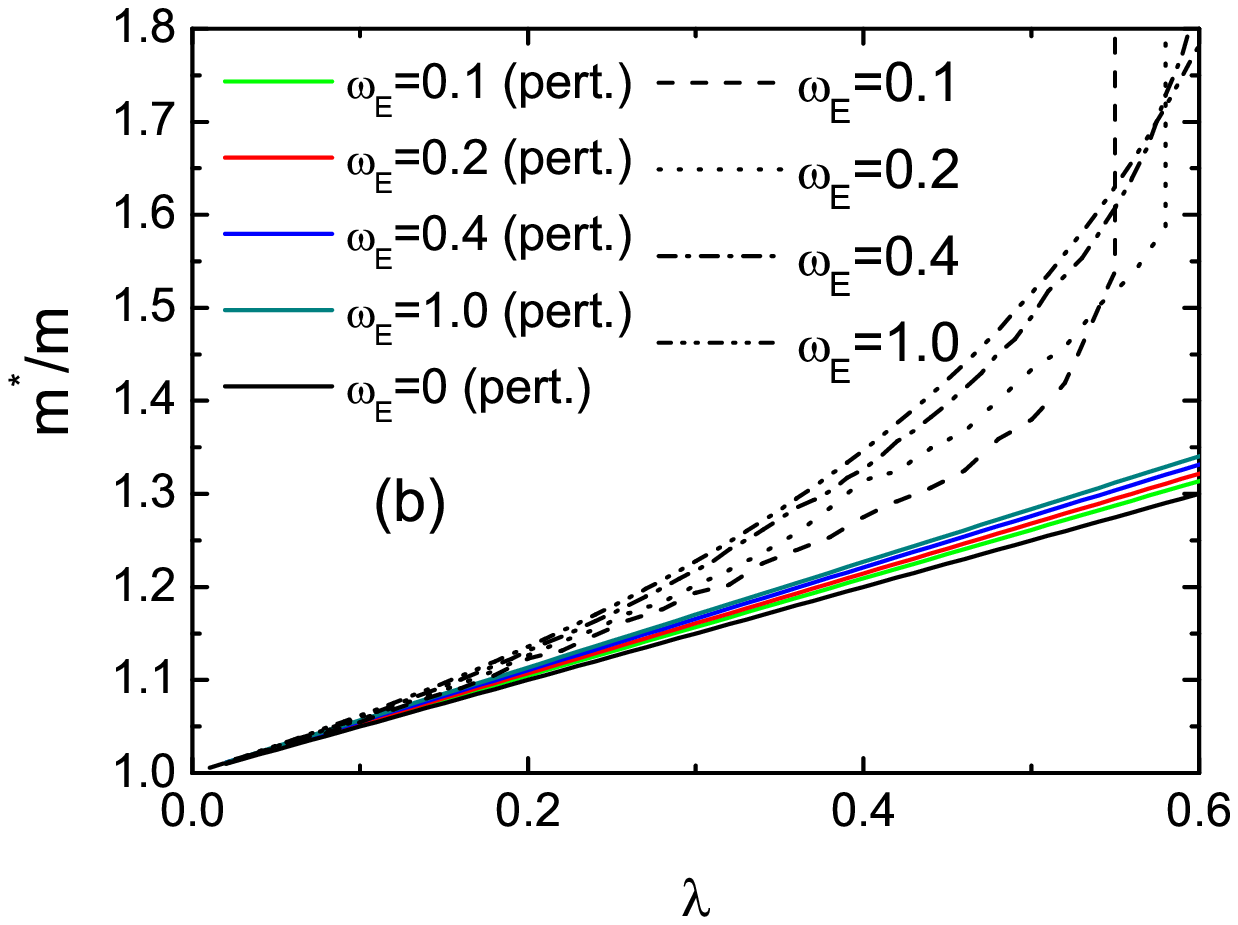}
\end{center}
\caption{ (color online) Effective mass $m^\ast/m$ vs. $\lambda$ for various phonon frequencies,
in two dimensions. Again, unlike the results in 1D in Fig. \protect\ref{fig2}, a `special' value of
$\lambda$ is clear --- $\lambda \approx 0.55$. However, for any non-zero
phonon frequency the behaviour below and above this special value is smoothly
connected. Only in the adiabatic limit does the behaviour change abruptly.\\
(b) Expansion of the weak coupling regime showing the numerical results alongside
the perturbation theory results. Agreement is not as good as in Fig. 3.}
\label{fig4}
\end{figure}

In Fig. \ref{fig2} we show the ground state energy $E_0$ vs. $\lambda$ for
various values of the phonon frequency; this is in one dimension. Fig. \ref%
{fig3} shows similar results in two dimensions. It is clear that as the
phonon frequency decreases, the crossover region near $\lambda \approx 1$
(actually, the `critical' value of $\lambda$, only valid in the adiabatic limit,
is closer to 0.55) becomes sharper. This is consistent with the result that, in the adiabatic
limit, there is a transition from a small polaron state to a free
electron-like state, in dimensions two and higher \cite{kabanov93}.
Nonetheless, as is known through other considerations \cite{lowen88}, for
any non-zero phonon frequency, the crossover is smooth.

To summarize this section, we have obtained numerically exact results for a
wide range of parameters, by using refinements to that used in Ref.
\onlinecite{trugman90,bonca99}. In particular, we obtain well converged
results over all coupling strengths \emph{and} for low phonon frequencies, $\omega_E << t$.
The results for low frequencies in particular illustrate a
rather abrupt crossover to a regime where multi-phonon processes are
prevalent. To what extent they play a crucial role even at intermediate
coupling strengths is the subject of the next section.

\section{perturbation theory}

Perturbation theory can be performed both from the weak and the strong
coupling limits.\cite{marsiglio95} In that work we obtained, to second
order (in $g$), in one dimension, the self energy
\begin{equation}
\Sigma_{\mathrm{1D}}(\omega+i\delta) = {\frac{ \lambda \omega_E \mathrm{sgn}%
(\omega - \omega_E) }{\sqrt{\biggl({\frac{\omega - \omega_E }{2t}}\biggr)^2
- 1}}};  \label{sig_weak_1D}
\end{equation}
which leads to a ground state energy:
\begin{equation}
E_0 = -2t\biggl( 1 + \lambda \sqrt{\frac{\omega_E }{4t + \omega_E}} \biggr).
\label{e0_weak_1D}
\end{equation}
This expression can be understood as follows: for very large frequency $\omega_E
>> t$ there is a correction by a factor $1+\lambda$ reminiscent of the mass
renormalization for the electron mass in a Fermi liquid state. On the other
hand, as the frequency becomes small the first order correction vanishes. In
fact, however, the most significant effect of the phonon coupling to a single electron
occurs for low phonon frequencies, while the effect disappears for high
phonon frequency. This is most readily seen by examining the quasi-particle
residue $z_0$ \cite{marsiglio95}
\begin{equation}
z_0 = 1 - {\frac{\lambda }{2}} \sqrt{\frac{t }{\omega_E}} {\frac{1 + {\frac{%
\omega_E }{2t}} }{(1 + {\frac{\omega_E }{4t}})^{3/2}}}.
\label{residue_weak_1D}
\end{equation}
or the effective mass, defined as
\begin{equation}
m/m^\ast = {\frac{1 }{2t}} {\frac{\partial^2 E(k) }{\partial k^2}}|_{k=0}.
\label{mass}
\end{equation}
For a momentum independent self energy (as in the second order weak coupling
expansion) these are simply related: $m^\ast/m = 1/z_0$. The residue clearly
approaches the non-interacting value, unity, as $\omega_E \rightarrow \infty$,
while it diverges as $\omega_E \rightarrow 0$. This indicates a breakdown in
(weak coupling) perturbation theory in this limit, which is consistent with the
fact established in Ref. [\onlinecite{kabanov93}] that the electron is polaron-like
for all coupling strengths, i.e. there is an abrupt change in character at $g=0$. In fact,
as established in Ref. [\onlinecite{holstein59}] for a two-site model, and in
Ref. [\onlinecite{kabanov93}], the effective mass diverges in the adiabatic limit for all
coupling strengths (in 1D), a limit which we now approach numerically in Fig. 2b.

In two dimensions, as mentioned in Section II, we use $\lambda = g^2\omega_E /(2\pi t)$.
This actually uses the electron density of states at the bottom of the band,
$N(-4t) = 1/(4\pi t)$, instead of the average density of states that is
commonly used, $N_{\mathrm{ave}}(0) = 1/(8t)$. The reason for this choice is
that we are studying the one electron sector, so the most pertinent density
of states is the one at the bottom.

The self energy (in two dimensions (2D)) in weak coupling is given by
\begin{equation}
\Sigma _{\mathrm{2D}}(\omega+i\delta) = {\frac{ \lambda }{2}}{\frac{%
8t\omega_E }{\omega - \omega_E}} K\Biggl[ \biggl({\frac{4t }{\omega -
\omega_E}}\biggr)^2\Biggr],  \label{sig_weak_2D}
\end{equation}
where $K(x) \equiv \int_0^{\pi/2} d\theta {\frac{1 }{\sqrt{1 - x \sin^2{%
\theta}}}}$ is the complete Elliptic integral of the first kind. This leads
to a ground state energy, which, in weak coupling, is:
\begin{equation}
E_0 = -4t \Biggl( 1 + {\frac{\lambda }{4}}{\frac{\omega_E }{t}}{\frac{1 }{1
+ \omega_E/(4t)}} K\Biggl[ {\frac{1 }{\bigl(1 + \omega_E/(4t)\bigr)^2}} %
\Biggr] \Biggr).  \label{e0_weak_2D}
\end{equation}
We can take the derivative of Eq. (\ref{sig_weak_2D}) to obtain:
\begin{equation}
m^\ast/m = 1 + {\frac{\lambda }{2}} {\frac{1 }{1 + \omega_E/(8t)}} E\biggl[{1 \over
(1+\omega_E/(4t))^2}\biggr],  \label{mass_weak_2D}
\end{equation}
where $E(x) \equiv \int_0^{\pi/2} d\theta \sqrt{1 - x \sin^2{\theta}}$ is
the complete Elliptic integral of the second kind. We have used, ${\frac{%
\partial K(x)}{\partial x}} = {\frac{1 }{2x}}\biggl({\frac{E(x) }{1 - x}} -
K(x)\biggr)$.

More familiar expressions are available, for cases when the arguments of the
complete elliptic integrals are close to unity. This occurs for $\omega_E <<
t$. Using $\lim_{m \rightarrow 1} K(m) = {\frac{1 }{2}} \mathrm{ln}(16/m_1)$, \cite{abramowitz72}
where $m_1 = 1 - m$, an approximate form for the ground state energy is:
\begin{equation}
E_0 \approx -4t \Biggl( 1 + {\frac{\lambda }{4}}{\frac{\omega_E }{t}}{\frac{%
1 }{1 + \omega_E/(4t)}} \mathrm{ln} \Biggl[ 4\sqrt{\frac{2t }{\omega_E}}{%
\frac{1 + \omega_E/(4t) }{\sqrt{1 + \omega_E/(8t)}}} \Biggr] \Biggr).
\label{e0_weak_2D_app}
\end{equation}
From Eq. (\ref{mass_weak_2D}) we obtain:
\begin{equation}
m^\ast/m = 1 + {\frac{\lambda }{2}} +\lambda {\frac{\omega_E }{16 t}}
\mathrm{ln}\biggl({\frac{32 t }{\omega_E}}\biggr) +O\Biggl(\Bigl({\frac{%
\omega_E }{t}}\Bigr)^2 \mathrm{ln}\biggl({\frac{t }{\omega_E}}\biggr)\Biggr).
\label{mass_weak_2D_app}
\end{equation}
Note that as $\omega_E/t \rightarrow 0$, the ground state energy approaches
the non-interacting value, while the result for the effective mass approaches the one
derived in the continuum limit by Cappelluti et al. \cite{cappelluti07}, and
the mass enhancement is half that expected when $E_F$ is large.

In Fig.3a, we show the ground state energy of the 2D Holstein model as a
function of $\lambda$ for a variety of phonon frequencies; we also show the result in the
adiabatic limit as $\omega _{E}/t\rightarrow 0$. For the latter case, we
adopted the iterative method described in Ref. [\onlinecite{kabanov93,marsiglio95}], and
used Lanczos  diagonalization for the electronic portion.
The abrupt transition occurs because we do not assume Bloch's theorem, and
translational invariance is broken for sufficiently strong coupling.
For non-zero phonon frequency we note the trend that
as $\omega_{E} \rightarrow 0$, the crossover from free-electron-like behaviour to
polaronic behaviour becomes more abrupt, though it is always smooth.\cite{lowen88}.

In Fig. 3b we show an expanded region in the weak coupling regime, where the
perturbation theory results are also plotted. Note that they are quite accurate for
all frequencies shown.

In Fig. 4a, we show the electron effective mass for the same parameters as in Fig. 3.
In strong coupling the effective mass grows rapidly with coupling strength, as shown.
However, this increase is even more pronounced as the phonon frequency decreases, until,
as the adiabatic limit is approached, the increase becomes very nearly abrupt above a
`critical' coupling strength, as determined through the adiabatic calculation.
At weak coupling, the effective mass is unity for large $\omega_E$ (not shown). As
the phonon frequency decreases, the effective mass grows; however, for smaller
phonon frequencies the effective mass will decrease again as the phonon frequency
decreases (as can be seen from the cases shown).
Both of these trends conspire to make the crossover more abrupt as the phonon frequency
approaches zero.

In Fig. 4b we show an expanded region in the weak coupling regime (no log scale), where the
perturbation theory results are also plotted. The results are certainly not as accurate as the
ground state energy; however, the inversion with phonon frequency noted above is clearly obtained.
%
%

\section{Mean phonon number and dispersion anomalies}

\begin{figure}[tp]
\begin{center}
\includegraphics[height=2.5in,width=2.5in]{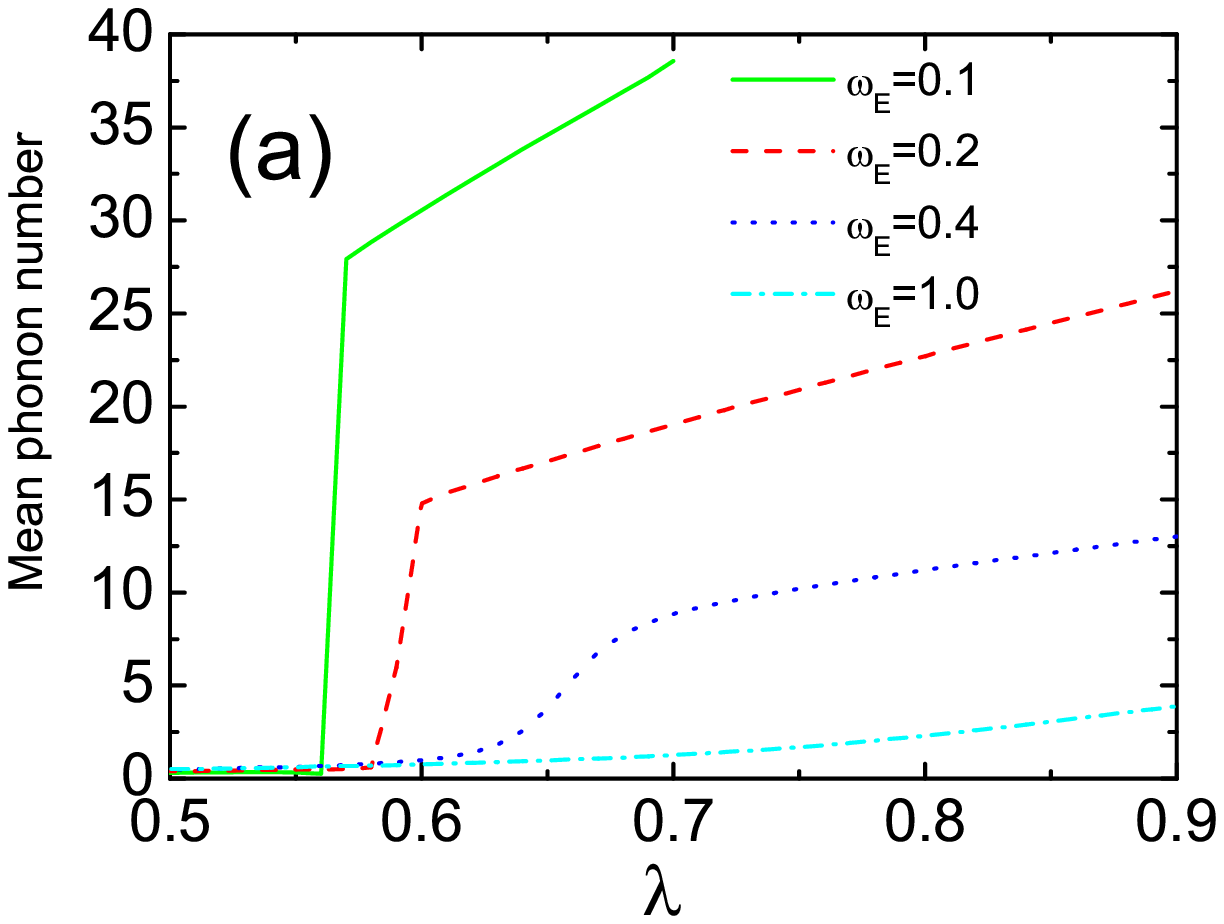}
\includegraphics[height=2.5in,width=2.5in]{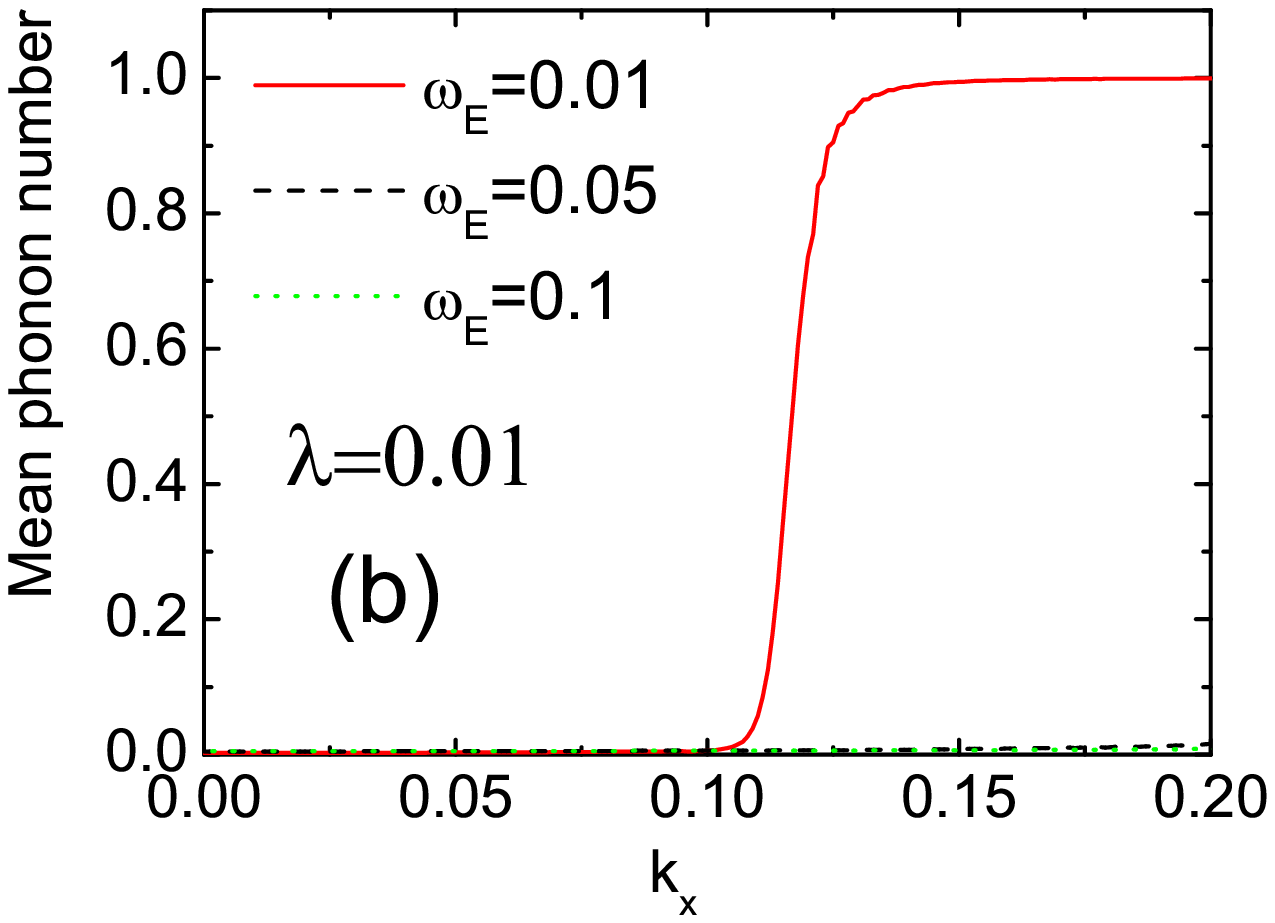}
\includegraphics[height=2.5in,width=2.5in]{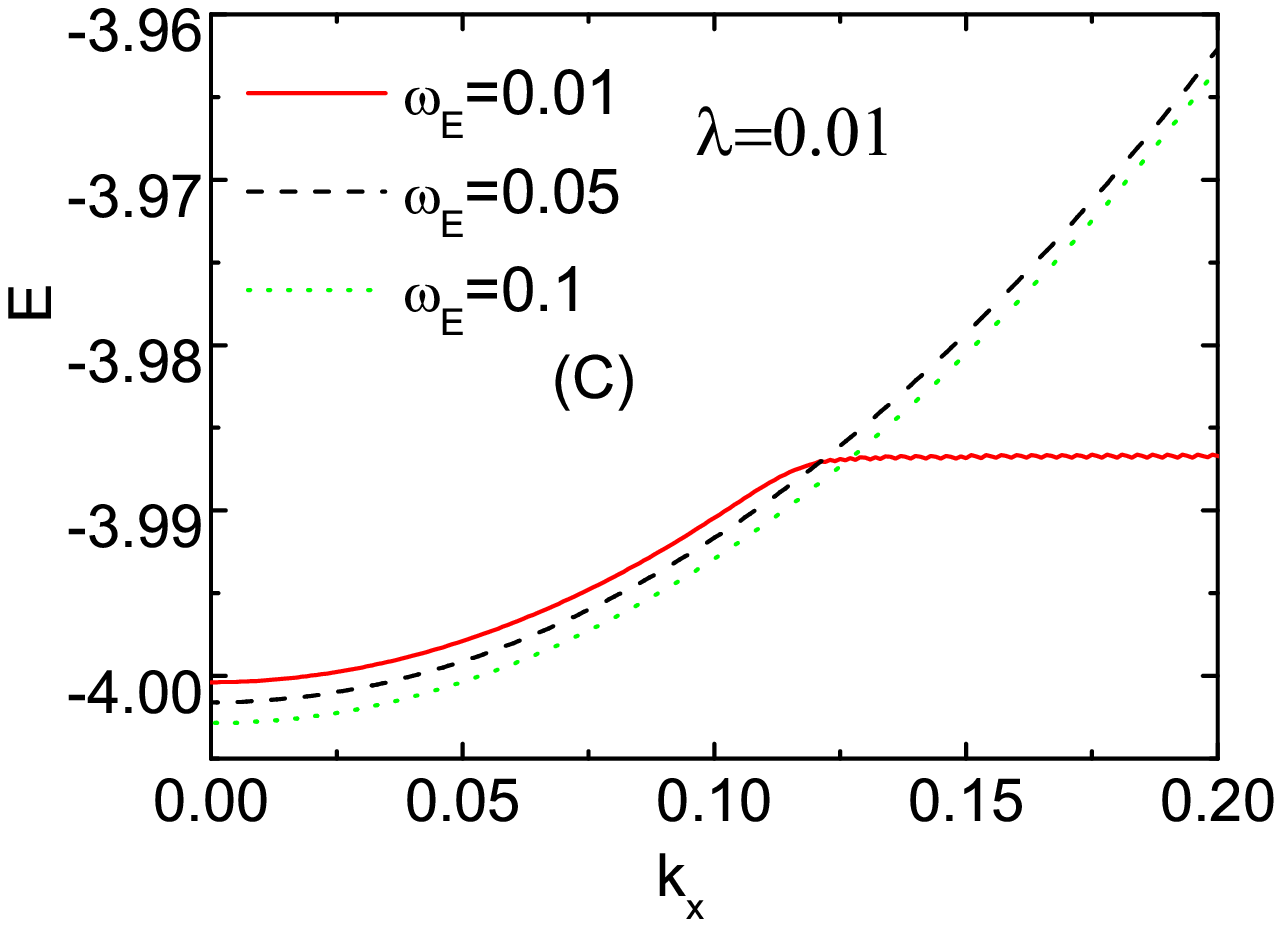}
\end{center}
\caption{ (color online) (a) Mean phonon number vs. coupling strength for various
phonon frequencies (in 2D). Note the increasingly abrupt behaviour as the phonon frequency decreases.\\
(b) Mean phonon number as a function of wave vector. Even for very small
coupling strength there is an abrupt increase when the phonon frequency is small
enough. Explanation is provided in the text, and is confirmed by (c) where the
energy as a function of wave vector is plotted for the same parameters as in (b).}
\label{fig5}
\end{figure}

Finally, we briefly examine the expectation of the number of phonons in
the ground state, and the impact on the electronic dispersion relation.
Restricting ourselves to two dimensions, we plot, in Fig.5a, the mean phonon number vs. coupling strength in the
intermediate coupling regime for several phonon frequencies. The same
trend as seen in Fig. 4 is apparent --- beyond a `special' coupling strength
the mean phonon number grows very abruptly from near zero to some value, $N_{s}$, after which
it continues to grow gradually as the coupling strength increases. The actual value of
$N_{s}$ is close to the central value of the Poisson distribution as predicted by strong
coupling perturbation theory \cite{marsiglio95}.

In Fig. 5b, we shows numerical results of the mean phonon
number as a function of total momentum $k_{x}$, of the electron-phonon system, for a few values of
phonon frequency, and for a very low value of coupling (so that the results are well converged).
Despite this small value of coupling, convergence is difficult because we use $\omega_{E}$ as small
as $0.01t$. We apply our self-adaptive Lanczos method by first converging the results for some high
momentum (say, $k_{x}$=0.3 --- we keep $k_y=0$), and then lower the value of $k_{x}$ in small increments,
and converge the calculation at each step, until we finally reach the desired end-point ($k_{x}=0$).
As Fig. 5b illustrates, for sufficiently small phonon frequency, the
mean phonon number shows a sharp increase from close to zero to nearly unity at
some wave-vector, say $k_{c}$. The reason for this is that the energy difference with the
ground state will eventually exceed a value of order $\omega_E$; at this point it becomes
energetically more favourable to use the zero momentum state (with much lower energy), and
simply excite a phonon with the required momentum. Confirmation of this explanation is
provided in Fig. 5(c), where the dispersion flattens abruptly beyond $k_c$, when the energy
exceeds that of the ground state by an amount approximately equal to $\omega_E$. It retains this value
because phonon momenta of any value are available with the same energy.

\section{Summary}

We have implemented an adaptation to the variational method first suggested by Trugman, specifically
to handle the adiabatic regime. In strong coupling our starting point leads to immediate convergence,
while in the intermediate coupling regime a `stepping-down' procedure allows for good convergence.
Even in weak coupling, if the phonon frequency is significantly lower than the hopping parameter, our
adaptive method is helpful, if not necessary.

By determining ground state properties as a function of decreasing phonon frequency we have
established a connection between numerical results at small but non-zero phonon frequency, and
adiabatic limit results obtained by using a semi-classical iterative procedure. It is clear that
in one dimension no weak coupling perturbation regime exists, while in two dimensions (and higher)
a definite weak coupling regime exists, and results derived within perturbation theory agree well
with numerical results down to very low frequencies. Finally, as the phonon frequency decreases, more
and more phonons are present in the ground state wave function, and these lead to anomalies in the electron
dispersion relation.

It is somewhat ironic that the most significant polaronic effects occur in the adiabatic regime, as $\omega_E/t \rightarrow 0$. This is where weak coupling perturbation theory breaks down completely. The second order result,
which is simply the so-called non-crossing approximation, fails to capture the rapid onset of multi-phonon excitations that form an integral part of the ground state wave function, as exemplified, for example, in Eq. (\ref{sc_zero}); this is a breakdown that, for example, Alexandrov \cite{alexandrov01} has repeatedly emphasized. At
the same time, the so-called Migdal approximation \cite{migdal58}, so key to the Eliashberg theory of superconductivity, is valid only in this limit. One then requires an understanding of how polaronic effects become minimized as more and more electrons are included in the problem. Apparently Pauli blocking plays an important role in mitigating the multi-phonon processes that constitute a single polaron. Future work will attempt to investigate this cross-over from
polaron to weak coupling behaviour.

\begin{acknowledgments}
This work was supported in part by the Natural Sciences and Engineering
Research Council of Canada (NSERC), by ICORE (Alberta), by Alberta Ingenuity, and by the Canadian
Institute for Advanced Research (CIfAR). We thank Stuart Trugman for helpful correspondence in the early
part of this work. FM is grateful to the Aspen Center
for Physics, where some of this work was done (summer, 2007).
\end{acknowledgments}

$^1$ present address: Dept. of Oncology, University of Alberta, Edmonton, AB, Canada T6G 1Z2 \\
$^2$ present address: Dept. of Mathematics, University of British Columbia, Vancouver, BC, Canada \\


\begin{thebibliography}{99}
\bibitem{comment1} In the implementation of Trugman's method the use of a
Bloch wave is implicit; nonetheless we include it explicitly in Eq. (\ref{sc_zero}). Furthermore,
in this representation we utilize phonon number states to reconstruct the exponential up to some
high cutoff.
\end{thebibliography}
\end{document}